# The Nonequilibrium Thermodynamics of Small Systems

Carlos Bustamante[*,@,&], Jan Liphardt[@], Felix Ritort

[*]Howard Hughes Medical Institute, University of California, Berkeley, CA 94720
[@]Departments of Physics University of California, Berkeley CA 94720, USA
[&]Department of Molecular and Cell Biology, University of California, Berkeley CA 94720, USA
[#]Departament de Física Fonamental, Facultat de Física, Universitat de Barcelona, Diagonal 647, 08028 Barcelona, Spain

*Introduction*

In the last several years, there has been an increased interest in small systems. Such systems are found throughout physics, biology, and chemistry and examples of them range from magnetic domains in ferromagnets, which are typically smaller than about 300 nm, to the biological molecular machines ranging in size from 2 - 100 nm, and the solid-like clusters important in the relaxation of glassy systems, with dimensions of a few nanometers. These systems manifest striking properties that are a direct result of their small dimensions and understanding these novel, size-dependent properties is of great current interest. For example, scientists are beginning to investigate the dynamics of the small biological motors that are responsible for converting chemical energy into useful forms of work in the cell. These systems operate away from equilibrium, dissipating energy continuously and making transitions between steady states. How is it that these tiny machines operate at energies only marginally above the thermal energy of the surroundings? How do the large fluctuations experienced by these systems affect their operation? Can we learn how to endow artificial nano-devices with similar properties? Can we formulate a nonequilibrium steady state thermodynamic description of these devices?

Until recently, scientists have lacked experimental methods to investigate the properties of small systems, such as the heat and work exchanged with their environment. This situation has changed as the result of the development of modern techniques of microscopic manipulation. In parallel, several theoretical results, collectively known as Fluctuation Theorems (FTs), have been derived and some of them have even been experimentally tested. The much-improved experimental access to the energy fluctuations of small systems, and the formulation of the principles that govern these exchanges and their statistical excursions, may ultimately serve as the basis towards the development of a theory of nonequilibrium thermodynamics of small systems. In this article, we will review some of these developments.

*Classical Thermodynamics and Properties of Small Systems*

Thermodynamics, a scientific discipline of the 19$^{th}$ century, was invented in an effort to rationalize the behavior of thermal machines that transformed heat into useful mechanical work (1). Thermodynamics was developed in the wake of the great success of classical mechanics in the 18$^{th}$ century. In many respects, however, thermodynamics clashed with its distinguished predecessor. For instance, in an effort to correspond with the corpuscular theory of light developed by Newton, the founders of thermodynamics fought to identify heat as a substance, the *caloric*, which was exchanged between bodies, never to be created or destroyed. It was only after Joule established the mechanical equivalent of heat, that the idea of the caloric was finally abandoned. Thermodynamics is a conceptual and abstract discipline built on two fundamental laws that have withstood the passing of time. The 1$^{st}$ law states that energy can be produced in two exchangeable forms, heat and work, and that the variation of energy content in a body is the sum of the heat supplied to the body and the work performed upon it. Energy is, thus, a conserved quantity. The 2$^{nd}$ law establishes a fundamental limitation on the amount of heat that can be transformed into work (2). Whereas work can be fully converted in to heat, the contrary is not true. The amount of heat that can be transformed into work is limited by the change of a state function, called entropy, between the initial and final states of the system. Moreover, for an isolated system undergoing a spontaneous process, the entropy must attain a maximum value. An important corollary of the 2$^{nd}$ law is the statement by Clausius that the total entropy of the universe always increases as energy progressively degrades into heat. Experimental tests of these laws quickly demonstrated their power and usefulness and thermodynamics soon became a new pier of science. At the end of the 19$^{th}$ century, spurred by the emerging molecular and atomic theory, as well as by difficulties in accommodating new observations, Boltzmann and others established the statistical mechanical foundations of thermodynamics as we know them today.

Thermodynamics describes the energy exchange processes of macroscopic systems: liquids, magnets, superconductors, and even black holes, fully comply with its laws. In macroscopic systems the observed behavior is reproducible and fluctuations (deviations from the typically observed or average behavior) are small. It is only in some special conditions that thermal fluctuations can produce readily detectable consequences: the opalescence of light in a fluid at the critical point is a prominent example. The importance of thermal fluctuations is determined by the law of large numbers. Let us first discuss fluctuations in equilibrium systems (3).

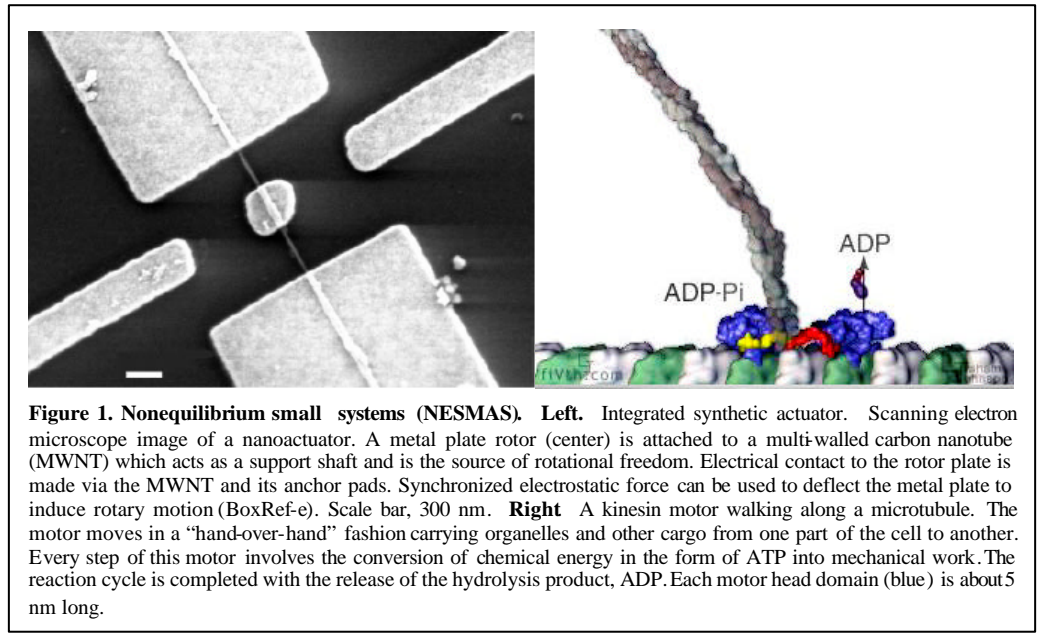

**Figure 1. Nonequilibrium small systems (NESMAS). Left.** Integrated synthetic actuator. Scanning electron microscope image of a nanoactuator. A metal plate rotor (center) is attached to a multi-walled carbon nanotube (MWNT) which acts as a support shaft and is the source of rotational freedom. Electrical contact to the rotor plate is made via the MWNT and its anchor pads. Synchronized electrostatic force can be used to deflect the metal plate to induce rotary motion (BoxRef-e). Scale bar, 300 nm. **Right** A kinesin motor walking along a microtubule. The motor moves in a "hand-over-hand" fashion carrying organelles and other cargo from one part of the cell to another. Every step of this motor involves the conversion of chemical energy in the form of ATP into mechanical work. The reaction cycle is completed with the release of the hydrolysis product, ADP. Each motor head domain (blue) is about 5 nm long.

In an ideal gas containing $N$ particles, the velocity distribution of each individual particle is described by the Maxwell distribution. The average total energy of the macroscopic system is a random quantity that is Gaussian distributed with average $(3/2)Nk_BT$ and variance $(3/2)(k_BT)^2N$. Therefore, the size of the fluctuations in the total energy, relative to its average value, are of order $N^{1/2}$, i.e. comparable for an individual particle ($N = 1$) but extremely small in macroscopic samples (where the number of particles is of the order of Avogadro's number, ~10$^{23}$). This simple argument shows the importance of large fluctuations for small systems when $N\sim O(1)$ or when the total energy of the system is few times the fundamental energy unit $k_BT$. In such small systems, thermal fluctuations can lead to observable large deviations from their average behavior and, therefore, these systems are not well described by classical macroscopic thermodynamics.

Systems of this type abound in nanotechnology, where motors with dimensions of less than 100 nm are being built (Figure 1, left), and inside the cell, where the biological function and efficiency of many molecular machines such as kinesin (Figure 1, right) is determined by their size. Kinesin is a highly processive molecular motor. In the cell, kinesin motors carry subcellular cargoes along microtubules. On average, a kinesin motor takes one 8 nm step (4) every 10 to 15 milliseconds. The conversion of chemical energy is tightly coupled to force generation and movement: one ATP molecule is hydrolyzed per step. How much energy does kinesin dissipate as it moves along the microtubule track? By subtracting the work done by the enzyme per step, 12 $k_BT$, from the chemical energy released by ATP hydrolysis (about 20 $k_BT$), we obtain an estimate of this machine's efficiency of about 60%. Thus, kinesin dissipates about 650 $k_BT$ per second into its environment.

Which class of physical systems, in the present context, can we refer to as small? In Figure 1 we show a diagram including different types of thermal systems classified according to their typical dissipation energy rate. As we can see, most of small systems are characterized by length-scales in the nanometer range.

Kinesin is one of many molecular machines. In the cell, these machines use the energy of bond hydrolysis to perform useful work, such as DNA replication, transcription, translation and repair. Molecular machines are unlike macroscopic machines, however, in that their small size may allow them to harness thermal fluctuations and rectify them using energy from chemical sources. For example, let us consider RNA polymerase, an enzyme that translocates along the DNA to produce a newly synthesized RNA strand (transcription). Although it has not yet been proven unequivocally, it is possible that during transcription the polymerase extracts energy from the bath to move, while it uses bond hydrolysis to insure that only "forward" fluctuations are captured, i.e., to rectify thermal fluctuations. The amount of energy required for the translocation step, the shape of the enzyme, the structural roughness and the information encoded in the steps along the track (in this case the base sequence of the DNA helix), are essential aspects attributable to the smallness of the system that ultimately determine its dynamics.

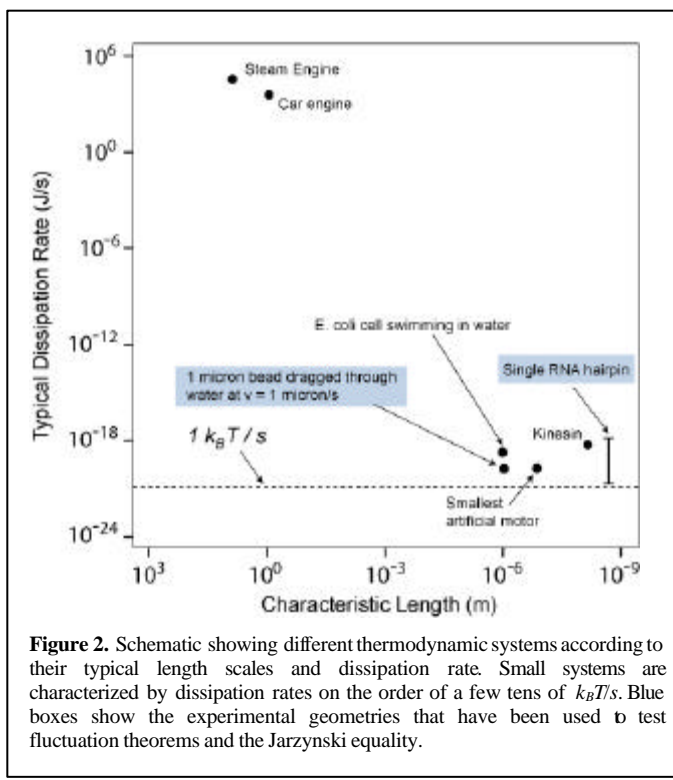

**Figure 2.** Schematic showing different thermodynamic systems according to their typical length scales and dissipation rate. Small systems are characterized by dissipation rates on the order of a few tens of $k_B T/s$. Blue boxes show the experimental geometries that have been used to test fluctuation theorems and the Jarzynski equality.

*Equilibrium and nonequilibrium systems.*

Just as in the case of macroscopic systems, when describing small systems we can distinguish between two situations in which the behavior and properties of a system do not change with time: equilibrium states and nonequilibrium steady states. Nonequilibrium steady states are characterized by net currents that flow across the system and whose macroscopic properties do not show observable time variation (e.g. a heat flux across a system in contact with two thermal sources, an electric current passing through a resistance, the motion of the kinesin along a microtubule. A constant input of energy is required to maintain such system in its present state. Because net energy is constantly dissipated by these systems, they operate away from equilibrium. Most biological systems such as the molecular machines described above and even whole cells belong to this category. Finally, in a non-steady state system, the most general case, one or more of the system's properties change in time.

The importance of nonequilibrium small systems (hereafter referred to as NESMAS) in the context of statistical physics was emphasized by Maxwell who "imagined a hypothetical being of intelligence, but molecular order of size, to illustrate limitations of the second law of thermodynamics" (5). The demon of Maxwell was "endowed with free will, and fine enough tactile and perceptive organization to give him the faculty of observing and influencing individual molecules of matter". According to Maxwell's paradox, a small demon sitting in the middle of a wall dividing a gas container could separate fast from slow molecules by opening a frictionless gate every time it sees a fast molecule moving towards the "fast" part. The demon would thus be capable of separating fast from slow molecules, leading to a net decrease in the total entropy of the system (without performing any mechanical work) and therefore violating the second law of thermodynamics.

Since it was proposed in 1871, the paradox has been studied by many scientists during the last century, and the demon has been exorcized several times. The ultimate solution to the paradox came in the 1960's in work by R. Landauer and C. H. Bennett (6) who showed that any decision made by the demon (i.e. allowing a given molecule to pass or not) entails gathering information from the system, information that has to be continuously erased and updated to make a new decision. This erasure saves the second law, as these authors showed that the erasure of one bit of information produces typically an amount of energy of the order of $k_BT$.

Two specific features of the demon envisaged by Maxwell call our attention, however: i) the demon is a genuine nonequilibrium system (as it must continuously replace old with new information); ii) the demon is small (a macroscopic Maxwell demon would not be suitable for molecular recognition as it could not influence the trajectory of the individual molecules). These two pivotal aspects, nonequilibrium and smallness, call for some general considerations that we unfold below.

*What are the properties of small systems?*

A key concept when describing fluctuations in small systems is the *control parameter*. The control parameter is the variable that must be specified to unambiguously define the system's state. Control parameters are thus variables that are fixed in a system while the other variables are allowed to fluctuate (See box 1).

An equilibrium state can be fully described by a small number of variables such as pressure and temperature. It is considerably more difficult to describe a nonequilibrium state. Micromanipulation technology, by giving direct access to control parameters of single macroscopic systems (see Box 1 and Figure 3), has opened up new opportunities to study NESMAS. By varying such parameters, one can perform controlled experiments in which the system is driven away from its initial state of equilibrium, and its response is then observed (7).

Let us focus on the example of the tethered polymer that is initially at equilibrium but that is then driven out of equilibrium by the action of an external perturbation (see Box 1). For example, a new end-to-end distance could be imposed on the polymer by moving the two walls further apart (Fig. 3, lower left). In that situation, the control parameter would be the distance between the two walls, and the nonequilibrium protocol would be fully specified by giving the wall-to-wall distance as a function of time $x(t) = X(t)/L$.

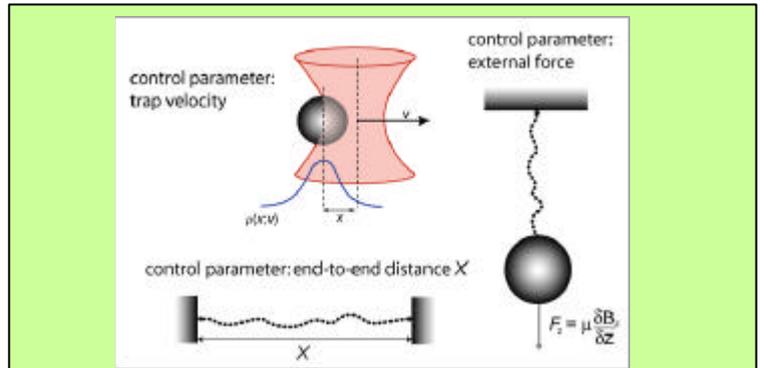

**Box 1 and Figure 3 The control parameter.** For small systems, the *control parameter* plays the role of the external variables (such as temperature, pressure, volume, mass, chemical potential) used to specify the different ensembles in statistical mechanics. In statistical mechanics all ensembles yield the same equation of state in the large volume limit, also called thermodynamic limit. Similarly, for small systems, the equation of state and the spectrum of fluctuations are fully determined by the choice of the control parameter. The figure illustrates small systems either in nonequilibrium steady-states or in equilibrium steady-states. **Bottom left and top right. Trapped polymers of contour length L.** The polymers, represented by a string of monomer units, are immersed in water at temperature $T$, hence atoms and bead continuously jiggle due to collisions with water molecules. The continuous formation and dissociations of hydrogen bonds between the monomers and water molecules produces a large friction between the polymer and water. The set of positions of all atoms or molecules in the polymer is specified by three-dimensional vectors $\{x_i\}$ which define the internal configuration of the system whereas $X$ specifies the end-to-end extension of the polymer. Under such conditions the force acting on the bead is a function of the relative extension $x = X/L$ and the total energy of the system, $U(\{x_i\},x)$, is of the order of $Nk_BT$, $N$ being the number of monomer units in the polymer. Two different equilibrium states can now be specified depending on which variable is taken as the control parameter (BoxRef-a). **Constant force**. For instance, we could fix the force acting on the bead by considering a magnetic bead with magnetic moment equal to **m** and applying a uniform external magnetic field gradient. By changing the value of the magnetic field gradient we could then control the force $F$ acting on the bead. Under such conditions, and for a fixed value of $F$, the relative extension $x$ is a fluctuating variable. The equation of state corresponding to this system is $F = g_F(\langle x \rangle)$ where $g_F$ is a one-variable function and $\langle x \rangle$ is the average relative fractional extension. For small forces we have $F = k\langle x \rangle$ where $k$ is a stiffness constant of the polymer at short extensions. This is the so-called linear or Hookean regime. **Constant extension**. Here, we fix the position of the bead at an extension equal to $X$ and we measure the force acting on the bead. The force $F$ will now be a fluctuating variable and the equation of state will have the form, $\langle F \rangle = g_x(x)$, where $g_x$ is a one-variable function and $\langle F \rangle$ is the average force. For small extensions we have $\langle F \rangle = kx$. Although for small deformations we obtain the same Hookean equation of state, in general $g_F \neq g_x$, meaning that the two equations of state are different for arbitrary extensions. In the large $L$ limit (the thermodynamic limit in this case) both equations of state coincide as expected and $g_F = g_x$ (BoxRef-a). In practice, however, both variables $X$ and $F$ fluctuate, and this effect must be taken into account when analyzing the force-extension trajectories (BoxRef-b). **Top left. An optically trapped bead**. Control parameters can arise in nonequilibrium states as well. This experimental geometry most closely approximates the situation considered by the Fluctuation Theorems. By translating the buffer-filled enclosure with respect to the trap, or steering the trap, a drag force is generated, which pushes the bead out of the potential minimum. In this geometry, the control parameter is the trap velocity $v$. In a steady state, the bead will on average lag a distance $x = v\gamma / k$ behind the center of the trap, where $\gamma$ is the drag coefficient and $k$ is the trap stiffness. Transitions between different steady-states can be produced by changing the trap velocity.

Since the system is small and is placed in a thermal bath, its dynamics will be random and each trajectory followed by the system will be different upon repetition of the same nonequilibrium protocol. Each trajectory can be represented by the time evolution of the positions of all atoms $\{x_i(t)\}$. Upon variation of the control parameter $x$, the variation of the total energy of the system $U(\{x_i\},x)$ has two contributions,

$$dU = \sum_i \left(\frac{\partial U}{\partial x_i}\right)_x dx_i + \left(\frac{\partial U}{\partial x}\right)_{\{x_i\}} dx = dQ + dW. \quad (1)$$

The first term corresponds to the variation of the energy as a result of the change of the internal configuration (that we interpret as heat), and the second term is the variation of the energy as a result of the perturbation applied by changing the value of the control parameter (that we interpret as work). If the control parameter changes from 0 to $x_f$, the total work done on the system is given by $W = \int_0^{x_f} dW = \int_0^{x_f} F dx$, and the heat exchanged in this nonequilibrium process is given by $Q = \Delta U - W$, where $\Delta U$ is the variation of the internal energy.

Stochasticity and fluctuations dominate the thermal behavior in small systems. Since the force $F = (\partial U/\partial x)_{\{x_i\}}$, is a fluctuating quantity, $W$, $Q$, $\Delta U$ will fluctuate as well and the amount of heat or work exchanged with the bath will fluctuate in magnitude and even sign. For a given nonequilibrium process we can define the work and heat probability distributions $P(W)$, $P(Q)$ as the histograms of the work and heat collected over an infinite number of repetitions of the same nonequilibrium protocol (to simplify notation without risk of confusion, we use the same letter $P$ for both distributions). In general, these distributions will directly depend on the experimental nonequilibrium protocol. The knowledge of such distributions is important to understand the details about how the system behaves and responds when subjected to a particular experimental process.

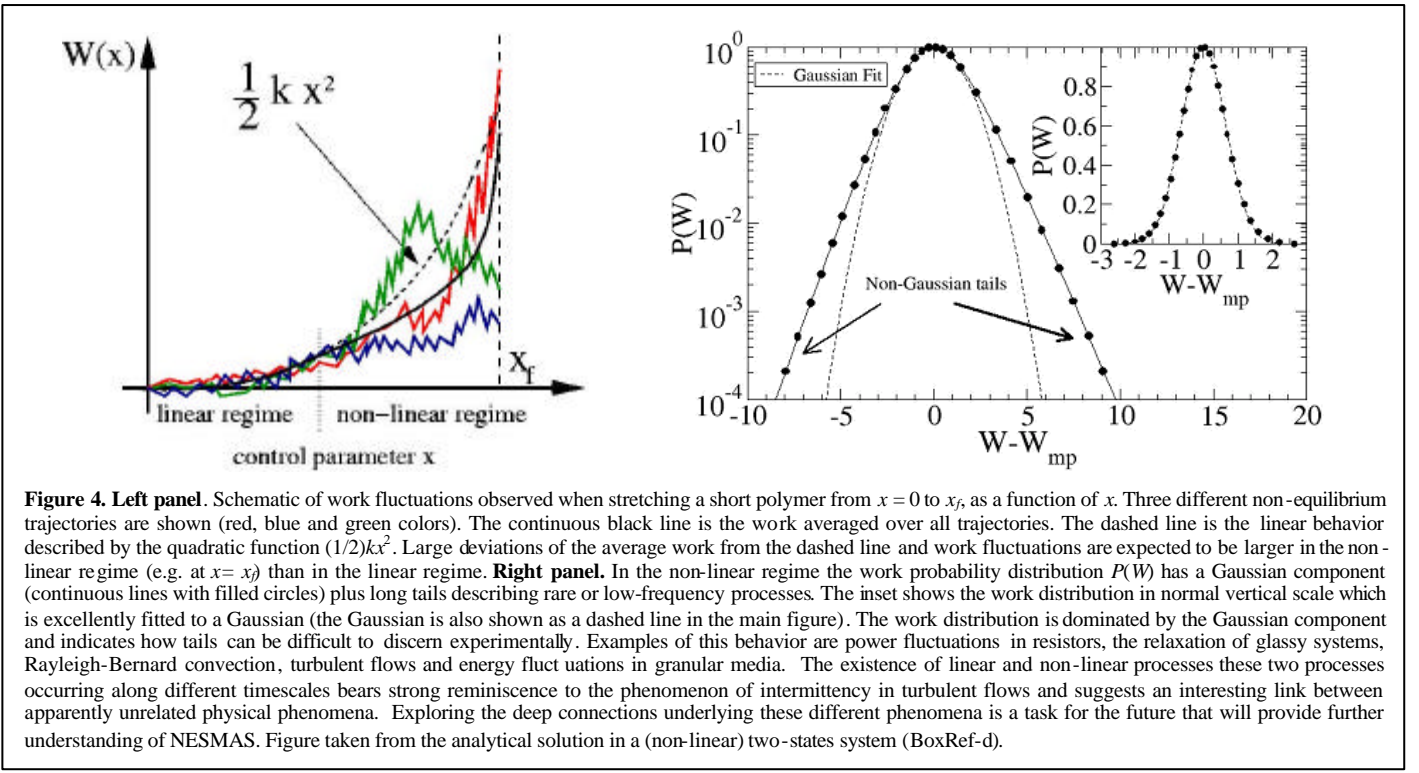

**Figure 4. Left panel**. Schematic of work fluctuations observed when stretching a short polymer from $x = 0$ to $x_f$, as a function of $x$. Three different non-equilibrium trajectories are shown (red, blue and green colors). The continuous black line is the work averaged over all trajectories. The dashed line is the linear behavior described by the quadratic function $(1/2)kx^2$. Large deviations of the average work from the dashed line and work fluctuations are expected to be larger in the non-linear regime (e.g. at $x = x_f$) than in the linear regime. **Right panel.** In the non-linear regime the work probability distribution $P(W)$ has a Gaussian component (continuous lines with filled circles) plus long tails describing rare or low-frequency processes. The inset shows the work distribution in normal vertical scale which is excellently fitted to a Gaussian (the Gaussian is also shown as a dashed line in the main figure). The work distribution is dominated by the Gaussian component and indicates how tails can be difficult to discern experimentally. Examples of this behavior are power fluctuations in resistors, the relaxation of glassy systems, Rayleigh-Bernard convection, turbulent flows and energy fluctuations in granular media. The existence of linear and non-linear processes these two processes occurring along different timescales bears strong reminiscence to the phenomenon of intermittency in turbulent flows and suggests an interesting link between apparently unrelated physical phenomena. Exploring the deep connections underlying these different phenomena is a task for the future that will provide further understanding of NESMAS. Figure taken from the analytical solution in a (non-linear) two-states system (BoxRef-d).

What type of distributions for heat and work can we expect in NESMAS? Lars Onsager proposed in 1931 that the regressions toward equilibrium, after applying a small external perturbation, could be treated as spontaneous fluctuations in the system (8). In NESMAS, fluctuations do not occur around an equilibrium state but around the most probable trajectory (see Figure 4). The heat and work produced along the nonequilibrium process reveals a pattern of fluctuations richer than that observed in equilibrium systems. When the perturbation is sufficiently small two sectors of heat and work fluctuations have been identified from the study of exactly solvable models (9,10). One sector corresponds to small deviations of the work/heat with respect to the most probable nonequilibrium trajectory. These are characterized by the so-called linear susceptibility (in the case of the tethered polymer, the value of the local stiffness constant $k=\partial F/\partial x$). Small deviations of the work/heat are frequent, Gaussian distributed, and satisfy the fluctuation-dissipation relation. The fluctuation dissipation relation states that the fluctuations of a system are related to the dissipative properties of the system when it is subject to an external interaction. The other sector is characterized by large deviations with respect to the most probable nonequilibrium trajectory. Large work/heat deviations are rare, their distributions typically show long exponential tails (Fig 4. Right panel), and these fluctuations do not satisfy the fluctuation-dissipation relation. The conjunction of this fast (Gaussian) and slow (exponential) processes has been observed in several areas of condensed matter physics. Examples are power fluctuations in resistors (11), the relaxation of glassy systems (12), Rayleigh-Bernard convection (13), turbulent flows (14) and energy fluctuations in granular media (15). The existence of these two processes occurring along different timescales bears strong reminiscence to the phenomenon of intermittency in turbulent flows. In such flows energy is dissipated in a series of cascade processes spanning widely separated length scales and time scales. This cascade scenario gives rise to the observed intermittent fluctuations in the velocity field of the fluid which show up as rare (low frequency) and large deviations of the velocity signal superimposed to a Gaussian distributed background. The presence of similar intermittent fluctuations in the energy exchanged of NESMAS suggests an interesting link between apparently unrelated physical phenomena. Exploring the deep connections underlying these different phenomena is a task for the future that will provide further understanding of NESMAS.

*Theoretical background: The Fluctuation Theorems (FTs)*

Nonequilibrium systems are characterized by irreversible heat losses between the system and the environment (generically called here "the thermal bath"). Recent developments towards a unified treatment of arbitrarily large fluctuations in small systems are embodied in fluctuation theorems, FTs, which relate the probabilities of a system exchanging a given amount of energy with the thermal bath in a nonequilibrium process (16).

In equilibrated systems, no net heat is transferred from the system to the bath, and therefore the probability of absorbing or releasing a given amount of heat must be identical, and the ratio $P(Q)/P(-Q)$ equals $1$[1]. This ratio becomes different from 1 under nonequilibrium conditions. Two important FTs were introduced by Evans and Searles for systems evolving from equilibrium toward a nonequilibrium steady state (17), and Gallavoti and Cohen for steady-state systems (18). These were based on numerical evidence obtained previously (19). In steady-state systems, heat is continuously produced by an external agent and transferred to the bath. The average amount of heat $<Q>$ so produced implies an increase in the total entropy of the system plus the environment equal to $<S>=<Q>/T$. The rate at which the system exchanges heat with the bath is called entropy production, $s=Q/Tt$ where $t$ is the interval of time over which the system exchanges the amount of heat $Q$. An explicit mathematical expression for the ratio $P(s)/P(-s)$ in steady states has been established under quite general conditions [2],

$$\lim_{t\to\infty}\frac{k_B}{t}\log\left(\frac{P_t(s)}{P_t(-s)}\right)=s. \qquad (2)$$

---

[1] Strictly speaking such relation holds only if the Hamiltonian is time-reversal invariant (no magnetic fields around). In the more general case we would just require $\int_0^\infty P(Q)dQ = \int_{-\infty}^0 P(Q)dQ$.

[2] In the original formulation by Gallavotti and Cohen, the quantity $s$ stands for the so-called *phase-space compression factor*. The phase space compression factor is related (but it is not necessarily identical) to the dissipated heat. The validity of Eq. (2) with $s$ identified with the rate of heat production has been shown to hold for $s\leq 1$ in the case of a bead dragged through water (Ref. 9, see below). For $s\geq 1$ the expression in the rhs of Eq. (2) is a more complicated function of $s$.

This expression indicates that in steady-state systems the heat delivered from the system to the bath (positive $s$) is more probable than heat absorbed by the system in the same conditions (negative $s$). Nonequilibrium steady-state systems always dissipate heat on average. For macroscopic systems the heat is an extensive quantity and therefore the ratio of probabilities $P_t(s)/P_t(-s)$ is exponentially large with the system size meaning that the probability of heat adsorption by macroscopic systems is insignificant. Our bodies, for example, are maintained in a nonequilibrium state by metabolic processes that dissipate heat all the time. For small systems such as molecular motors that translocate along a molecular track, however, the probability of absorbing heat can be significant. Although on average molecular motors produce heat, it is possible, as mentioned above, that they move by rectifying thermal fluctuations, corresponding to occasionally capturing heat from the thermal bath.

who derived the corresponding FT for the case of a system, initially in thermal equilibrium, that is driven out of equilibrium by the action of an external agent (BoxRef-c). Let $x_F(s)$ denote a nonequilibrium process that starts at equilibrium in state A, lasts for a time $t$ and ends in a nonequilibrium state at B. We will call this the forward process. Let us now consider the reverse process were the system starts at equilibrium in state B and ends in a nonequilibrium state at A. The nonequilibrium protocol for the reverse process $x_R(s)$ is time-reversed respect to the forward one, $x_R(s) = x_F(t-s)$ so both processes last for the same time $t$. Let $P_F(W)$ and $P_R(-W)$ stand for the work probability distributions along the forward and reversed processes respectively. In such conditions the Crooks FT (CFT) applies,

$$\frac{P_F(W)}{P_R(-W)} = \exp\left(\frac{W - \Delta G}{k_B T}\right) \quad (4)$$

i.e., the probability to dissipate a given amount of work along the forward process is larger than the probability to adsorb the same amount of work along the reverse process. The CFT shares much resemblance with the Gallavotti-Cohen FT in the context of steady state systems (Eq. 2) if we make the identification $\sigma t = W_{dis}/T$ with $W_{dis} = W - \Delta G$. The main difference lies in the asymptotic validity of (1) as compared to the general validity of (Eq. 4) at all finite t. The origin of such difference arises from the fact that in (Eq. 4) we assume the system is initially in thermodynamic equilibrium in state A (along the forward process) and state B (along the reverse process) while in (1) we just consider that the system has reached a nonequilibrium steady state. The JE (Eq. 3) is a particular result of the CFT if we just rewrite (Eq.4) in the form $P_R(-W) = P_F(W)\exp(-(W-\Delta G)/k_B T)$ and integrate it from $W = -\infty$ to $W = +\infty$. The left hand side gives 1 (probability distributions are normalized) while the r.h.s. gives the r.h.s. of (Eq. 3).

The CFT relates hysteresis along the forward and reverse nonequilibrium processes. If a rubber band is deformed along a hysteresis cycle, the mechanical work exerted upon the band during the stretching part of the cycle (the area below the force-deformation curve) is always larger than the corresponding area for the release part of the cycle. The area enclosed inside the force deformation curve corresponding to the cycle is equal to the amount of mechanical work that is dissipated in the form of heat by the rubber band. In contrast to macroscopic systems, in small systems stochastic fluctuations can be very large and sometimes the stretching curve lies below the releasing curve showing that FTs can be useful to predict (and quantify) under which conditions large deviations from the average or macroscopic behavior are experimentally measurable. One of the most interesting consequences of Eq. 4 is that it can be used to predict free-energy differences by measuring the value of the work at which the forward and reverse work distributions cross each other. From Eq. 4 $P_R(-W) = P_F(W)$ if $W = \Delta G$. That is, the value of the work at which both distributions cross must not depend on the nonequilibrium protocol and is equal to the average reversible work done on the system and in turn equal to the free energy change involved in the process. For Gaussian work distributions it can be shown that the crossing value of the work, $\Delta G$, is just the mean of the average work along the forward and reverse processes (Fig. 4 left panel). In Figure 4 (right panel) we show work distributions obtained by pulling an RNA hairpin at three pulling speeds (2, 7 and 20 pN/s). The two vertical lines indicate the region of work values where the crossing is observed. As predicted by the CFT, the crossing is independent of the pulling speed (BoxRef-f).

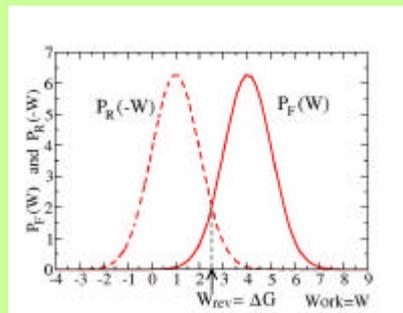 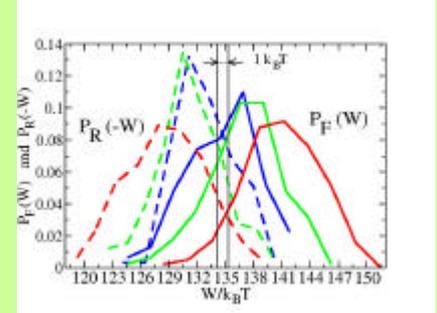

FTs shed light on Loschmidt's paradox. In 1876, Loschmidt (20) raised an objection to Boltzmann's derivation of the second law from Newton's laws of motion. According to Loschmidt, since the microscopic laws of mechanics are invariant under time reversal, there must also exist entropy-decreasing evolutions, in apparent violation of the second law. The FTs show how macroscopic irreversibility arises from time reversible microscopic equations of motion: time-reversed trajectories *do* occur, but with increasing system size they become vanishingly rare, and the second law in its conventional form emerges. In his celebrated monograph "What is life?" (21), Schroedinger emphasized the importance of inverted motion saying: "The true physical picture includes the possibility that even a regularly going clock should all at once invert its motion and, working backward, rewind its own spring –at the expense of the heat of the environment-."

At least 6 other FTs have been reported, and differ by the details of the system's dynamics (stochastic vs. deterministic), nature of the thermostat, and initial conditions (equilibrium or nonequilibrium steady state) (22). A parallel development began in 1997 with the report of a nonequilibrium work relation by C. Jarzynski (23), called the Jarzynski equality (hereafter referred to as JE). Consider a system kept in contact with a bath at temperature $T$ whose equilibrium state is determined by the value of the control parameter $x$. Initially, the system is at equilibrium and the control parameter is $x_A$. The nonequilibrium process is obtained by changing $x$ according to a given protocol $x(t)$, from $x(0) = x_A$ to $x(t) = x_B$. The JE states that for any nonequilibrium process that starts at the equilibrium state (A) and ends at the final state (B),

$$\exp\left(-\frac{\Delta G}{k_B T}\right) = \left\langle \exp\left(-\frac{W}{k_B T}\right) \right\rangle \quad (3)$$

where $\Delta G$ is the free-energy difference between the equilibrium states A and B, and where the average $\langle \ldots \rangle$ is taken over an infinite number of repeated nonequilibrium experiments carried out with the protocol $x(t)$. Frequently, the JE is recast in the form $\langle \exp(-W_{dis}/k_B T) \rangle = 1$ where $W_{dis} = W - \Delta G$ is the dissipated work along a given trajectory.

The exponential average appearing in the JE implies that $\langle W \rangle = \Delta G$ or $\langle W_{dis} \rangle = 0$ which is the statement of the 2nd law of thermodynamics in terms of free energy and work (1). An important consequence of the JE is that, although on average $W_{dis} = 0$, there must always exist nonequilibrium trajectories with $W_{dis} = 0$ for the equality to hold. These trajectories, sometimes referred to as transient violations of the second law, stand for large fluctuations in the work that ensure that the microscopic equations of motion are time-reversal invariant. The JE is remarkable, for it implies that it should be possible to determine the free energy difference between initial and final equilibrium states of the process from nonequilibrium realizations of this process (24). As shown below, this result is of great practical importance.

G. E. Crooks was subsequently able to relate various FTs by deriving a generalized FT for stochastic microscopically reversible dynamics (25) (see Box 2). Such consolidation has proceeded, and is now understood that neither the details of the thermostat (26), nor the somewhat differing interpretations of the extensive property considered in Eq. 2 as entropy production, entropy production rate, dissipated work, exchanged heat etc. lead to fundamentally distinct FTs. All these theoretical advances have benefited greatly from advances in micromanipulation tools, which make it possible to measure energy fluctuations in NESMAS, directly test the validity of FTs, and scrutinize some fundamental assumptions of statistical mechanics.

*Computer simulations and the Fluctuation Theorems*

Computer simulations have played an essential role in the development of the FTs. Indeed, the first paper on the subject, the 1993 report by Evans, Cohen and Morriss (19), included molecular dynamics simulations of a two-dimensional gas of disks. As shown in Fig. 6, over suitably short times, spontaneous ordering of the gas was observed, in agreement with the expression they derived for the probability of fluctuations of the shear stress of a fluid in a nonequilibrium steady state.

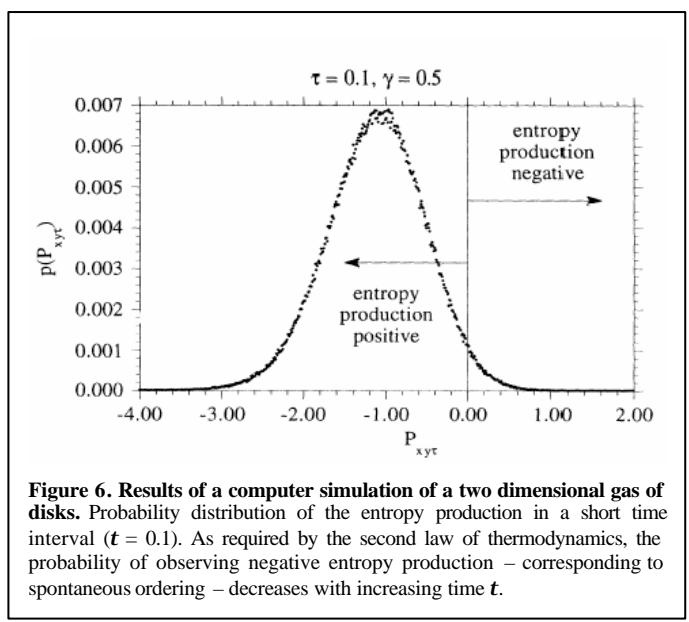

**Figure 6. Results of a computer simulation of a two dimensional gas of disks.** Probability distribution of the entropy production in a short time interval ($t = 0.1$). As required by the second law of thermodynamics, the probability of observing negative entropy production – corresponding to spontaneous ordering – decreases with increasing time $t$.

Computer simulations of nonequilibrium systems have continued to play an important role in the field, primarily due to the difficulty of setting up and characterizing suitably NESMAS. It is now also appreciated that FTs, and especially the JE, can potentially be used to improve the performance of molecular dynamics simulations. Since the JE asserts that free energy estimates can be obtained by Boltzmann-weighted averaging over numerous independent, irreversible simulations, and not only from a single slow growth, or adiabatic simulation, it should be possible to parallelize free-energy calculations over fast irreversible paths (23). Hendrix and Jarzynski compared the performance of the two approaches in 2001 (29). They found that reliable free energy estimates could indeed be obtained from independent, irreversible simulations, and that this "fast growth" approach permitted easy estimation of errors, unlike slow growth simulations. Fast and slow growth thermodynamics integration methods have been directly compared in calculations of the potential of mean force between two methane molecules. For a given total computational cost, the fast growth method was shown to provide comparable or better free energy estimates than conventional slow growth and umbrella sampling (30).

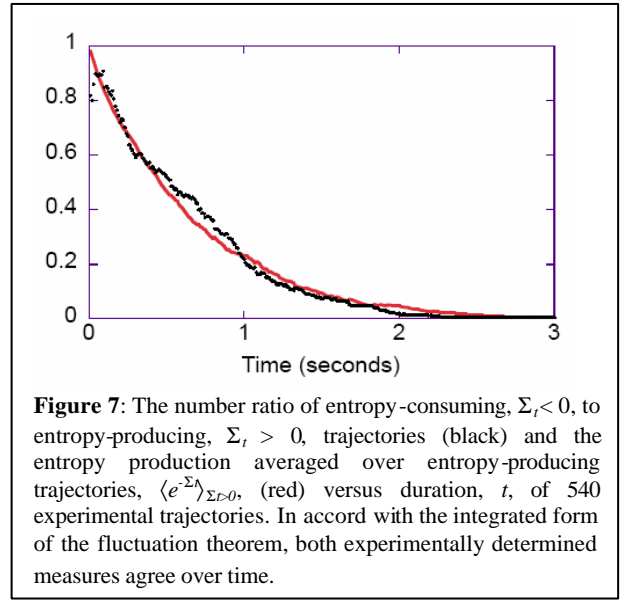

**Figure 7**: The number ratio of entropy-consuming, $\Sigma_t < 0$, to entropy-producing, $\Sigma_t > 0$, trajectories (black) and the entropy production averaged over entropy-producing trajectories, $\langle e^{-\Sigma_t} \rangle_{\Sigma_t > 0}$, (red) versus duration, $t$, of 540 experimental trajectories. In accord with the integrated form of the fluctuation theorem, both experimentally determined measures agree over time.

The JE is now also being used in steered molecular dynamics simulations, to efficiently calculate potentials of mean force (31). When the work distributions are Gaussian, as is the case when the system is steered by suitably stiff springs, a simplified form of the JE (its cumulant expansion to second order) can be used with little loss of accuracy and with faster convergence. In addition, it is also now appreciated that the accuracy of free energy estimates obtained from simulations and experiments can be significantly improved by taking both forward and reverse trajectories into account. Shirts and coworkers (32) derived a simple formula for the variance of free energy estimates using the Bennett acceptance ratio method (a method that optimizes and guarantees the lowest variance for the estimate value of the free-energy difference for general work distributions). Examples of direct relevance to biophysics include the direct measurement of the free-energy landscape in biomolecular processes (e.g. the free-energy encountered by the glycerol molecule dragged through the ion channel protein acquaglyceroporin GlpF (33))

## *Experimental tests of Fluctuation Theorems*

The development of experimental techniques that make it possible to manipulate small systems, and to measure their responses to external perturbations, has enabled researchers to directly test several of the FTs. The first experimental test of the Gallavotti-Cohen FT was done by Ciliberto and Laroche in 1998 (34) in the study of Rayleigh-Benard convection. In 2002, the group of Denis J. Evans in Australia (35) verified an integrated form of the Evans-Searles FT (17) by repeatedly dragging microscopic beads through water with an optical trap, and computing the entropy production for each bead trajectory.

The likelihood of entropy consuming trajectories relative to entropy producing trajectories was seen to be precisely what was predicted by theory. For small times (milliseconds), entropy consuming trajectories could be readily observed, and as expected, the classical "bulk" behavior was recovered for longer times (seconds). As shown in Fig. 7, the ratio of entropy-consuming to entropy-producing trajectories (black points) and the entropy production averaged over entropy-producing trajectories (red) agreed over time, in accord with the integrated FT.

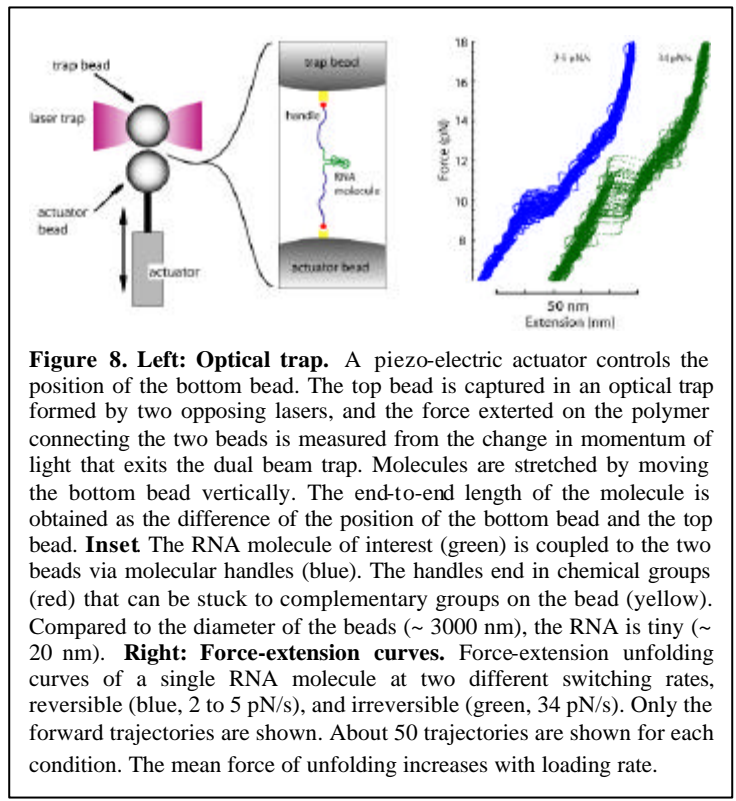

**Figure 8. Left: Optical trap.** A piezo-electric actuator controls the position of the bottom bead. The top bead is captured in an optical trap formed by two opposing lasers, and the force exerted on the polymer connecting the two beads is measured from the change in momentum of light that exits the dual beam trap. Molecules are stretched by moving the bottom bead vertically. The end-to-end length of the molecule is obtained as the difference of the position of the bottom bead and the top bead. **Inset**. The RNA molecule of interest (green) is coupled to the two beads via molecular handles (blue). The handles end in chemical groups (red) that can be stuck to complementary groups on the bead (yellow). Compared to the diameter of the beads (~ 3000 nm), the RNA is tiny (~ 20 nm). **Right: Force-extension curves.** Force-extension unfolding curves of a single RNA molecule at two different switching rates, reversible (blue, 2 to 5 pN/s), and irreversible (green, 34 pN/s). Only the forward trajectories are shown. About 50 trajectories are shown for each condition. The mean force of unfolding increases with loading rate.

In the same year, the JE was tested by mechanically stretching a single molecule of RNA, both reversibly and irreversibly, between its folded and unfolded conformations (36) (Figure 8). When the polymer was unfolded slowly, the average forward and reverse trajectories could be super-imposed, indicating a reversible reaction. When the polymer was unfolded more rapidly, the mean unfolding force increased and the mean refolding force decreased. The folding/unfolding cycle was thus hysteretic, indicating the dissipation of work.

The biophysical relevance of the JE was noted by Hummer and Szabo (37) who showed how equilibrium potentials of mean force could be constructed using single-molecule experiments carried out under nonequilibrium conditions. Application of the JE to the irreversible work trajectories recovered the free energy profile of the unfolding process to within $k_BT/2$ of its best independent estimate, the mean work of reversible stretching. The implementation and test of the JE provided an example of its use as a bridge between the statistical mechanics of equilibrium and nonequilibrium systems. This work also extended the thermodynamic analysis of single molecule manipulation data beyond the context of equilibrium experiments. Of course, the problem with this approach is that, for the equality to hold, the average has to be taken over an infinite number of trajectories, which is unrealizable in experiments or simulations. For a finite number of experiments the free-energy estimate we obtain by applying Eq. 3 is biased (31,38), and it is therefore essential to know extrapolation methods or error estimates in such conditions. The JE can be also seen as a consequence of the validity of a FT derived by Crooks (see Box 2).

Experimental tests of the FTs are in continuous progress. Technical improvements have recently enabled a direct test of a relation similar to Eq. 2, rather than the integrated form (39). Electrical circuits are also now being employed as the driven, dissipative system by Sergio Ciliberto and coworkers. Compared to the tests involving trapped beads or stretched polymers, electrical circuits are less prone to drift and other systematic biases, and permit much higher numbers of trajectories, enabling one to investigate systems with larger dissipation rates. In these experiments, an electrical dipole was maintained in a nonequilibrium steady state by injection of a constant current, and the probability distributions of work and heat were collected and shown to be in very good agreement with the FT applicable to this situation (11). By recording several hours of fluctuation data from their driven electrical dipole, Garnier and Ciliberto (11) were able to investigate the exchanged heat and work with unusually high resolution, and ultimately detect the non-Gaussian tails in the heat distribution predicted by Van Zon and Cohen (9).

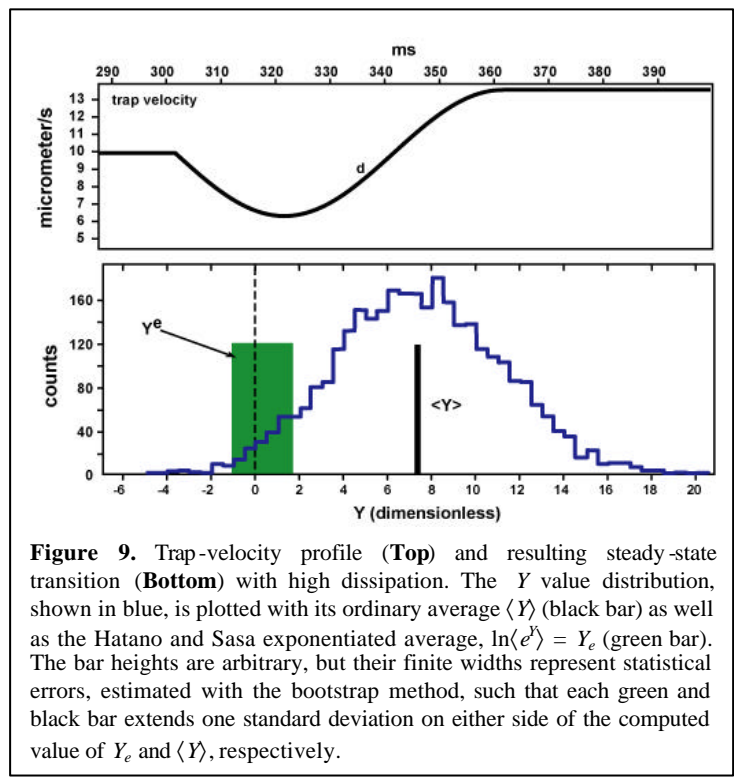

**Figure 9.** Trap-velocity profile (**Top**) and resulting steady-state transition (**Bottom**) with high dissipation. The $Y$ value distribution, shown in blue, is plotted with its ordinary average $\langle Y \rangle$ (black bar) as well as the Hatano and Sasa exponentiated average, $\ln\langle e^Y \rangle = Y_e$ (green bar). The bar heights are arbitrary, but their finite widths represent statistical errors, estimated with the bootstrap method, such that each green and black bar extends one standard deviation on either side of the computed value of $Y_e$ and $\langle Y \rangle$, respectively.

Understanding of the JE has also improved. One of the key assumptions used in its derivation is that the system starts at equilibrium, and is then driven out of equilibrium by some external influence. Many systems of interest, however, whether they are a biological molecular machine or a nanophotonics device, do not necessarily fulfill that assumption. Rather, they typically execute irreversible transitions between nonequilibrium steady states. In 1998, Oono and Paniconi (43) proposed a general phenomenological framework encompassing nonequilibrium steady states and transitions between such states. Three years later, Hatano and Sasa (44), building upon that work showed that the exponential average of a specific quantity $Y$ related to a dissipated work, should be equal to 1 for arbitrary transitions between nonequilibrium steady states, $\langle e^{-Y} \rangle = 1$ Several aspects of this prediction are of interest. First, $Y = 0$ when the process is carried out reversibly, suggesting that in the more general case the nonnegative value $\langle Y \rangle$ provides a measure of the irreversibility of the process. Second, for the system studied by Hatano and Sasa, $\langle Y \rangle = 0$ is equivalent to a generalized Clausius inequality proposed within Oono and Paniconi's phenomenological steady-state thermodynamic framework (43). This prediction has been tested by measuring the dissipation and fluctuations of microspheres optically driven through water (45) as shown schematically in Fig. 3 (top). Although the mean generalized dissipated work was always greater than zero (Fig. 9), $\ln\langle e^{-Y}\rangle$ was approximately equal to 0 for three different nonequilibrium systems, supporting a prediction of Hatano and Sasa's for arbitrary steady states and irreversible transitions between them.

The Crooks' Fluctuation Theorem (CFT, see Box 2) has also been tested, and used to obtain free energy estimates from nonequilibrium experiments (46). One of the most interesting consequences of the theorem is that the value at which the forward and reverse work distributions cross must not depend on the nonequilibrium protocol and is equal to the free-energy change involved in the process. In Figure 5 we show experimental work distributions obtained by pulling an RNA hairpin at three pulling speeds (2, 7 and 20pN/s). The two vertical lines indicate the region of work values where the crossing is observed. As predicted by the CFT, the crossing is independent of the pulling speed The CFT was verified under weak and strong nonequilibrium conditions by measuring the irreversible mechanical work during the unfolding and refolding of a small RNA molecule with optical tweezers. Once the CFT had been verified, it was used it to obtain folding free-energies for various biomolecules unfolded through nonequilibrium trajectories by determining the crossing between the unfolding and refolding work distributions. Compared to the earlier studies of the JE, in which the typical dissipated works were quite small (less than 4 $k_BT$), the CFT makes it possible to recover free energies from reactions with dissipations of up to 50 $k_BT$, and presumably corresponding to systems driven out of the linear response regime.

FTs predict a symmetry in the probability of absorbing and releasing a given amount of energy between the system and the environment. However they do not tell us anything about the mathematical form of the energy distributions. Most of the experiments carried out until now have investigated systems with energy distributions that are Gaussian to a very good approximation. The variance of such energy distributions is then proportional to the average dissipated energy, a result that is characteristic of nonequilibrium systems in the linear response regime. In general, NESMAS are expected to show more complicated energy distributions, where large and rare energy fluctuations contribute as long tails to a Gaussian distribution describing small and frequent deviations. Until now, our knowledge of the general properties of NESMAS energy distributions is quite limited as these phenomena are only slowly becoming accessible to experimental observation. Therefore we do not yet understand the relevance and meaning of these tails although we suspect that these might carry information about the NESMAS.

*Conclusion and Outlook*

The study of nonequilibrium small systems (NESMAS), including the building and characterization of small nanodevices, and the description of the many molecular machines in the cell, is becoming of central importance in physics, chemistry and biology. In particular, many processes in the cell are carried out by very few numbers of molecules in which nonequilibrium conditions predominate. Because they are intrinsically microscopic, these systems are subjected to large deviations and thus, many central cell processes, such as protein synthesis, energy generation, and catalysis are inherently noisy. That the cell somehow manages to coordinate these noisy processes is one of the remarkable, and still poorly understood, facts of complex biological systems. How is it that the cell is capable of coordinating all these processes in which the signals are essentially buried in noise is one of the remarkable facts of complex biological systems that are still not well-understood. The correct description of these and other small, nonequilibrium systems awaits the development of a generalized nonequilibrium thermodynamic formalism that will be able to relate thermodynamic or quasi-thermodynamic quantities with processes involving both steady-state and nonequilibrium states.

What makes us believe that such general theory can exist and, at the same time, is useful? There is an interesting connection between thermal processes in NESMAS and other problems in physics currently under examination. We have already mentioned the intermittency effects that are observed in turbulent flows and other systems in condensed-matter physics. These facts leads us to emphasize the importance of understanding NESMAS in order to unify the description of energy exchange processes that occur in many apparently unrelated systems. Whereas most of the emphasis in this paper has been on biological systems we foresee that most of the developments in this exciting area will get further confirmation and endorsement by studying physical systems that can be better controlled and where nonequilibrium experiments are more easily repeatable.

Before concluding, we want to point out the importance of FTs in quantum systems. These were not discussed in the present paper, yet there exist several quantum versions of the classical FTs in the literature. However, no quantum FT has yet been put under experimental scrutiny. Yet, such experiments might show interesting surprises. Quantum coherence might completely change the physical limits where large deviations could be observable and therefore where FTs can be tested. In classical systems the size represents a main limitation for the observation of large fluctuations. That limit could be tested in quantum systems under appropriate conditions. These as well as many other exciting problems remain a future challenge for scientists working with small systems, a fertile ground where physics, chemistry and biology converge.

*Figure sources:*

**Fig. 2a** A.M. Fennimore, T.D. Yuzvinsky, Wei-Qiang Han, M.S. Fuhrer, J. Cumings and A. Zettl, *Nature* **424**, 408-410 (2003).
**Fig. 2b** www.fivth.com, Graham Johnson Medical Media, graham@fivth.com
**Fig. 4** F. Ritort, *J. Stat. Mech.: Theor. Exp*. P10016 (2004).
**Fig. 6** D.J. Evans, E.G.D. Cohen, and G.P. Morriss, *Phys. Rev. Lett*. **71**, 2401 (1993).
**Fig. 7** D.J. Evans, E.G.D. Cohen, and G.P. Morriss, *Phys. Rev. Lett*. **71**, 2401 (1993).

*Box references:*

a) D. Keller, D. Swigon and C. Bustamante, *Biophys. J.,* **84**, 733 (2003).
b) J. M. Schurr, *Biophys. J*., **84**, 178A Part 2 Suppl. S (2003).
c) G. E. Crooks, *Phys. Rev. E* **60**, 2721 (1999).
d) F. Ritort, *J. Stat. Mech.: Theor. Exp*. P10016 (2004).
e) A.M. Fennimore, T.D. Yuzvinsky, Wei-Qiang Han, M.S. Fuhrer, J. Cumings and A. Zettl, *Nature* **424**, 408-410 (2003).
f) D. Collin, F. Ritort, C. Jarzynski, S. B. Smith, I. Tinoco Jr. and C. Bustamante, *Nature* **437**, 231-234 (2005)